\documentclass[a4paper,10pt,twoside]{cpc-hepnp}
\usepackage{CJK,upgreek,fancyhdr}
\usepackage{multicol}
\usepackage{graphicx}
\usepackage{booktabs}
\usepackage{amssymb,bm,mathrsfs,bbm,amscd}
\usepackage[tbtags]{amsmath}
\usepackage{lastpage}

\begin{document}
\begin{CJK*}{GB}{gbsn}

\fancyhead[c]{\small Submitted to Chinese Physics C}
\fancyfoot[C]{\small 010201-\thepage}


\title{Reevaluation of thermonuclear reaction rate of $^{50}$Fe($p$,$\gamma$)$^{51}$Co\thanks{Supported by Natural Science Foundation of
Inner Mongolia Autonomous Region of China (2013MS0916) and National Natural Science Foundation of China (11490562, 11405228)}}

\author{%
      Li-Ping Zhang$^{1}$ \quad Jian-Jun He$^{2;1)}$\email{hejianjun@nao.cas.cn} \quad Wan-Dong Chai$^{1}$ \quad Su-Qing Hou$^{3}$ \quad Li-Yong Zhang$^{3}$}
\maketitle

\address{%
$^1$ College of Physics and Electronic Information Engineering, Chifeng University, Chifeng 024000, China\\
$^2$Key Laboratory of Optical Astronomy, National Astronomical Observatories, Chinese Academy of Sciences, Beijing 100012, China\\
$^3$ Institute of Modern Physics, Chinese Academy of Sciences, Lanzhou 730000, China\\
}

\begin{abstract}
The thermonuclear rate of the $^{50}$Fe($p$,$\gamma$)$^{51}$Co reaction in the Type I X-ray bursts (XRBs) temperature range has been reevaluated based on a
recent precise mass measurement at CSRe lanzhou, where the proton separation energy $S_p$=142$\pm$77 keV has been determined firstly for the $^{51}$Co
nucleus. Comparing to the previous theoretical predictions, the experimental $S_p$ value has much smaller uncertainty. Based on the nuclear shell model
and mirror nuclear structure information, we have calculated two sets of thermonuclear rates for the $^{50}$Fe($p$,$\gamma$)$^{51}$Co reaction by utilizing
the experimental $S_p$ value. It shows that the statistical-model calculations are not ideally applicable for this reaction primarily because of the low
density of low-lying excited states in $^{51}$Co. In this work, we recommend that a set of new reaction rate based on the mirror structure of $^{51}$Cr
should be incorporated in the future astrophysical network calculations.
\end{abstract}



\footnotetext[0]{\hspace*{-3mm}\raisebox{0.3ex}{$\scriptstyle\copyright$}2013
Chinese Physical Society and the Institute of High Energy Physics
of the Chinese Academy of Sciences and the Institute
of Modern Physics of the Chinese Academy of Sciences and IOP Publishing Ltd}%

\begin{multicols}{2}

\section{Introduction}
Type I X-ray bursts (XRBs) arise from thermonuclear runaways on the accreted envelopes of neutron stars in close binary
systems~\cite{bib:woo76,bib:jos77}. During the thermonuclear runaway, an accreted envelope enriched in H and He may be transformed to matter strongly
enriched in heavier species (up to A$\sim$100~\cite{bib:sch01,bib:elo09}) via the rapid proton capture process
(rp-process)~\cite{bib:wal81,bib:sch98,bib:woo04}. Please see, e.g., Refs.~\cite{bib:lew93,bib:str06,bib:par13} for reviews on the XRBs.

The rp-process is largely characterized by localized $({\rm p},\gamma)$-$(\gamma,{\rm p})$ equilibrium within particular isotonic chains near the proton
drip-line. In such an equilibrium situation the abundance distribution within an isotonic chain depends exponentially on nuclear mass differences as the
abundance ratio between two neighboring isotones is proportional to $\exp[S_p/kT]$ ($S_p$: proton separation energy, $T$: temperature of the stellar
environment). In particular, those isotonic chains with sufficiently small $S_p$ values (relative to XRB temperatures - at 1 GK, $kT$$\approx$100 keV)
need to be known with a precision of at least 50$\sim$100 keV~\cite{bib:sch98,bib:par09}. In order to compare model predictions with observations of
the light curves~\cite{bib:sch06}, reliable nuclear-physics inputs, e.g., precise $S_p$ values and nuclear structure information, are needed for those nuclei
along the rp-process path.

Recently, precise mass measurements of nuclei along the rp-process path have become available. These measurements were made at the HIRFL-CSR
(Cooler-Storage Ring at the Heavy Ion Research Facility in Lanzhou)~\cite{bib:xia02} in an IMS (Isochronous Mass Spectrometry) mode. The proton
separation energy of $^{51}$Co has been experimentally determined to be $S_p^\mathrm{IMP}$=142$\pm$77 keV for the first time~\cite{bib:shu14}.
Although the estimated values in the previous Atomic Mass Evaluations (i.e., $S_p$=240$\pm$210 keV in AME85~\cite{bib:aud85}, 290$\pm$160 keV
in AME93~\cite{bib:aud93}, 90$\pm$160 keV in both AME95~\cite{bib:aud95} and AME03~\cite{bib:aud03}) agree with this experimental value
within 1~$\sigma$ uncertainty, the experimental value is significantly more precise.

Previously, the impact of $Q$-value (i.e., $S_p$ value) for the
$^{50}$Fe($p$,$\gamma$)$^{51}$Co reaction was studied~\cite{bib:par09} based on the old $S_p$($^{51}$Co) value of AME03. In the XRB `short' model,
it shows that the uncertainty of $Q$-value has very large impact on the final yields of $^{51}$Cr and $^{52}$Fe, whose yields can be significantly
affected by factors of 4.9 and 2.0, respectively, by changing $Q$ to $Q+\Delta Q$. Therefore, a precise $S_p$ ($^{51}$Co) value is very important for
constraining the final XRB yields. In this work, we have derived the thermonuclear rates of $^{50}$Fe($p$,$\gamma$)$^{51}$Co based on the new experimental
$S_p$ ($^{51}$Co) value, with which the resonant and direct capture (DC) rates are recalculated. This precise $S_p$ value allows the uncertainty in the rate of the
$^{50}$Fe($p$,$\gamma$)$^{51}$Co reaction to be dramatically reduced (e.g., see Ref.~\cite{bib:hjj14}), and hence the XRB yields can be well constrained too.

\section{Previous reaction rates}
\subsection{Available rates}
The thermonuclear rate of the $^{50}$Fe($p$,$\gamma$)$^{51}$Co reaction was firstly estimated by Van Wormer {\it et al.}~\cite{bib:wor94} based
entirely on the properties of 6 resonances in the mirror nucleus $^{51}$Cr, because of no experimental level structure of $^{51}$Co available. A value of
$S_p$=240 keV estimated from AME85 was utilized in that paper. This rate was estimated later~\citep{bib:rau00} with a statistical-model code using the
Hauser-Feshbach formalism (NON-SMOKER~\cite{bib:rau98}) based on the different $S_p$ values of $^{51}$Co predicted by the finite-range droplet
macroscopic model (FRDM)~\cite{bib:frdm} ($S_p$=1.369 MeV) and ETSFIQ mass model~\cite{bib:pea96} ($S_p$=1.659 MeV). Later on, this rate was
calculated by Fisker {\it et al.}~\cite{bib:fis01} under a framework of shell model with a value of $S_p$=90 keV from AME95. In addition, some theoretical
rates calculated using statistical models are available in JINA REACLIB\footnote{http://groups.nscl.msu.edu/jina/reaclib/db}~\cite{bib:cyb10}, which
were estimated by the statistical model with different $S_p$ values. The predicted rates differ from one another by up to several orders of magnitude over
the typical XRB temperatures. Moreover, the mirror $^{51}$Cr nucleus exhibits a low level-density structure near the proton threshold of $^{51}$Co, and
therefore the reliability of such statistical model calculations may be questionable.

Below, we refer to previously available rates using the nomenclature adopted in JINA REACLIB database. Explicitly, the \emph{laur} rate refers to exactly
the rate estimated by~\cite{bib:wor94}; the \emph{rath} rate adopts exactly the one calculated by~\cite{bib:rau00}. The \emph{rath, thra} rates are the
statistical-model calculations with FRDM and ETSFIQ masses, respectively. The recent \emph{ths8} rate is the theoretical one from
T. Rauscher~\cite{bib:cyb10}. In addition, the rate calculated by Fisker {\it et al.}~\cite{bib:fis01} was revised with a value of $S_p$=355 keV, and
referred to as \emph{nfis} in JINA REACLIB~\cite{bib:cyb10}.

\subsection{Existing problems}
By a careful survey, we find that there are several mistakes made in the previous paper of Fisker {\it et al.}~\cite{bib:fis01}. In general, the direct-capture
(DC) rate, which was expressed by their Eq.~(15), should be multiplied by a factor of 2. Concerning the $^{50}$Fe($p$,$\gamma$)$^{51}$Co reaction,
we have recalculated all the rates (DC, resonant and total ones) by utilizing the parameters listed in their Table I. (p. 261), and the comparison is made in
Table~\ref{tab1}. It can be seen that all rates listed in Table I (p. 286) of Ref.~\cite{bib:fis01} are significantly different from the present ones, and it
implies that some mistakes were made in the previous work. Unfortunately, we cannot not find the exact source of these errors.

\begin{center}
\tabcaption{ \label{tab1} Ratio of reaction rates between present and previous calculations~\cite{bib:fis01}.}
\footnotesize
\begin{tabular*}{85mm}{c@{\extracolsep{\fill}}ccccc}
\toprule & \multicolumn{4}{c}{Ratio (Present / Previous)}  \\
\cline{2-4}
T9 & DC rate$^a$ & Resonant rate$^b$  & Total rate \\
\hline
0.1   &	9.8E-02	&	3.4E+08	&	3.4E+08	\\
0.2   & 6.4E-02	&	4.6E+04	&	2.5E+04	\\
0.3   & 5.3E-02	&	1.1E+00	&	1.1E+00	\\
0.4   & 4.8E-02	&	1.3E+00	&	1.3E+00	\\
0.5   & 4.4E-02	&	1.4E+00	&	1.4E+00	\\
0.6   & 4.2E-02	&	1.3E+00	&	1.3E+00	\\
0.7   & 4.0E-02	&	1.3E+00	&	1.3E+00	\\
0.8   & 3.9E-02	&	1.2E+00	&	1.2E+00	\\
0.9   & 3.8E-02	&	1.2E+00	&	1.2E+00	\\
1.0   & 3.7E-02	&	1.1E+00	&	1.1E+00	\\
2.0   & 3.2E-02	&	8.9E-01	&	7.9E-01	\\
3.0   & 3.0E-02	&	9.4E-01	&	6.0E-01	\\
4.0   & 2.9E-02	&	9.8E-01	&	4.8E-01	\\
5.0   & 2.8E-02	&	9.9E-01	&	3.6E-01	\\
6.0   & 2.7E-02	&	9.9E-01	&	2.5E-01	\\
7.0   & 2.7E-02	&	9.9E-01	&	1.8E-01	\\
8.0   & 2.7E-02	&	9.9E-01	&	1.3E-01	\\
9.0   & 2.6E-02	&	9.9E-01	&	9.5E-02	\\
10.0  & 2.6E-02	&	9.9E-01	&	7.3E-02	\\
\bottomrule
\end{tabular*}
\vspace{0mm}

$^a$: Calculated by the analytical Eq.~(15) of Ref.~\cite{bib:fis01} with an enhanced factor of 2 as explained in the text.
$^b$: Calculated by the analytical Eq.~(7) of Ref.~\cite{bib:fis01} with exactly the same parameters listed in Table I.
\end{center}
\vspace{0mm}

In addition, Van Wormer {\it et al.}~\cite{bib:wor94} estimated the $^{50}$Fe($p$,$\gamma$)$^{51}$Co reaction rate relying on 6 resonances in the
mirror $^{51}$Cr, and neglected the DC contributions. They used a value of $S_p$=240 keV adopted from AME85. In fact, only two resonances, i.e., at
$E_r$=0.51, 0.92 MeV, dominate the total resonant rate; the former contributes the rate below 0.4 GK, while the latter overwhelmingly contributes in the
region of 0.4$\sim$2 GK. We have reproduced very well (less than 5\% deviation) the strength $\omega\gamma$ values for the five resonances (i.e.,
$E_r$=0.51, 0.53, 1.12, 1.32 and 1.65 MeV) listed in their Table~15 in Ref.~\cite{bib:wor94}, with an approximation of
$\omega\gamma \approx \omega\Gamma_\gamma$ (since $\Gamma_p \gg \Gamma_\gamma$). As for the `key' resonance at $E_r$=0.92 MeV
($J^{\pi}$=9/2$^-$), a value of $\omega\gamma$=3.8$\times$10$^{-2}$ eV was listed in Ref.~\cite{bib:wor94}. With the same approximation of
$\omega\gamma \approx \omega\Gamma_\gamma$, we get a strength value about 14\% larger than the previous one (we guess that Van Wormer
{\it et al.} made the same approximation). According to Eq.~(6) of Van Wormer {\it et al.}, a proton width value of $\Gamma_p$=2.08$\times$10$^{-6}$
is obtained for this high-spin 9/2$^-$ (with $\ell$=5 transfer) resonance, with a nuclear channel radius of $R=1.26\times(1+50^\frac{1}{3})$ fm and an
assumed spectroscopic factor of $C^2S$=0.1. Thus, the above approximation is invalid for this resonance, and its strength is calculated to be
1.04$\times$10$^{-5}$ eV. Now, this `key' 0.92-MeV resonance makes only negligible contribution to the total rate; the 0.51-MeV resonance dominates the
total rate below $\sim$1 GK, while the 1.12-MeV resonance dominates the rate in the temperature region of 1$\sim$2~GK.

\section{New reaction rates}
In this work, we will use the new experimental value of $S_p^\mathrm{IMP}$($^{51}$Co)=142$\pm$77 keV to recalculate the thermonuclear reaction rate
of $^{50}$Fe($p$,$\gamma$)$^{51}$Co. As for a typical ($p$,$\gamma$) reaction, the total thermonuclear reaction rate consists of the resonant and
DC rates of proton capture on ground state and all thermally excited states in the target nucleus weighted with their individual population
factors~\cite{bib:rol88}. The reaction rates for the $^{50}$Fe($p$,$\gamma$)$^{51}$Co reaction are calculated as described in the following subsections.

It is well-known that the population of an excited state $E_x$ relative to the ground state of the nucleus can be described by the Boltzmann
probability function~\cite{bib:rol88}:
\begin{eqnarray}
P(E_x)=\frac{2J_r+1}{2J_0+1}\times\mathrm{exp}\left(-\frac{E_x}{kT} \right),
\label{eq1}
\end{eqnarray}
where $J_0$ and $J_r$ are the spins of the ground state and excited resonant state, respectively. According to Eq.~(\ref{eq1}) the probabilities of populating
the first-excited state ($E_x$=764.9 keV) relative to the ground state in $^{50}$Fe are about 1.4$\times$10$^{-38}$ and 0.06 at 0.1 and 2 GK, respectively.
Therefore, contributions from these thermally populated excited states can be entirely neglected in the temperature region of XRB interested.

\subsection{Resonant rates}
In this work, the resonant rate is calculated by Eq.~(7) in Ref.~\cite{bib:fis01}, i.e., the well-known narrow resonance formalism~\cite{bib:wor94,bib:rol88},
\end{multicols}
\ruleup
\begin{equation}
\label{eq2}
N_A\langle \sigma v \rangle_\mathrm{res}=1.54 \times 10^{11} (AT_9)^{-3/2} \omega\gamma \mathrm{exp} \left (-\frac{11.605E_r}{T_9} \right) \quad [\mathrm{cm^3s^{-1}mol^{-1}}].
\end{equation}
\begin{multicols}{2}
\noindent
Here, the resonant energy $E_r$ and strength $\omega\gamma$ are in units of MeV. For the proton capture reaction, the reduced mass $A$ is defined by
$A_T$/(1+$A_T$) (here, target mass $A_T$=50 for $^{50}$Fe). The resonant strength $\omega\gamma$ is defined by (i.e. Eq.~(8) in Ref.~\cite{bib:fis01})
\begin{eqnarray}
\omega\gamma=\frac{2J+1}{2(2J_T+1)}\frac{\Gamma_p\times\Gamma_\gamma}{\Gamma_\mathrm{tot}}.
\label{eq3}
\end{eqnarray}
Here, $J_T$ and $J$ are the spins of the target and resonant state, respectively. $\Gamma_p$ is the partial width for the entrance channel,
and $\Gamma_\gamma$ is that for the exit channel.

Peak temperatures in recent hydrodynamic XRB models have approached 1.5--2~GK~\cite{bib:woo04,bib:jos10}. Here, we will consider the reaction rate
which holds for a temperature region up to 2.5~GK. For the $^{50}$Fe($p$,$\gamma$)$^{51}$Co reaction, a temperature of 2.5 GK corresponds to a
Gamow peak $E_r \approx$1.96 MeV with a width of $\Delta \approx$1.50 MeV~\cite{bib:rol88}. Thus, its resonant rate is determined by the resonances
with maximum energy up to $\sim$2.71 MeV. It can be seen that contributions from the resonances presented in the following
Tables~\ref{tab2}~\&~\ref{tab3} are sufficient to account for the resonant rate at XRB temperatures.

\subsubsection{Rate based on shell model}
In the previous shell-model calculation~\cite{bib:fis01}, all the resonant parameters are listed in Table I (see p.~261). Here, we need to recalculate the
resonant rate based on the new experimental value of $S_p^\mathrm{IMP}$($^{51}$Co)=142$\pm$77 keV. Actually only two quantities need to be
changed, i.e., resonance energy $E_r$ and proton width $\Gamma_p$ ($\Gamma_\gamma$ independent on $S_p$ value). Resonance energy can be
calculated easily by $E_r^\mathrm{Revised}=E_x^\mathrm{Fisker}-S_p^\mathrm{IMP}$. The proton width can be calculated by
$\Gamma_p^\mathrm{Revised}=\frac{P_\ell(E_r^\mathrm{Revised})}{P_\ell(E_r^\mathrm{Fisker})}\times\Gamma_p^\mathrm{Fisker}$.
Thereinto, the Coulomb penetrability factor $P_\ell$ can be calculated by the subroutine {\tt RCWFN}~\cite{bib:bar74}, with same optical-model
parameter (i.e., radius $R=1.26\times(1+50^{1/3})$ fm) as in Refs.~\cite{bib:fis01,bib:wor94}.

The resonant parameters are listed in Table~\ref{tab2}, and the revised resonant rates (referred to as Res$^{shell}$) in Table~\ref{tab4}. Our calculation
shows that the first two resonances at $E_x$=839.1, 866.7 keV dominate the total resonant rate up to $\sim$2 GK, beyond which other two resonances at
2076.8, 2597.1 keV begin to make significant contributions.

\begin{center}
\tabcaption{\label{tab2} Revised resonant parameters based on the previous work~\cite{bib:fis01}. Here, the excitation energy ($E_x^{th}$) and
resonance energy ($E_r$) are in units of keV.}
\footnotesize
\begin{tabular*}{85mm}{cccccc}
\toprule $E_x^{th}$ & $J^{\pi}$ & $E_r$  & $\Gamma_p$ (eV) & $\Gamma_\gamma$ (eV) & $\omega\gamma$ (eV)\\
\hline
\hphantom{0}839.1  & 5/2$^-$ & \hphantom{0}697.1  &	1.14E-05 & 2.05E-05	& 2.19E-05 \\
\hphantom{0}866.7  & 7/2$^-$ & \hphantom{0}724.7  &	4.93E-06 & 5.08E-05	& 1.80E-05 \\
1720.1 & 5/2$^-$ & 1578.1 &	9.64E-05 & 6.33E-03	& 2.85E-04 \\
1857.7 & 3/2$^-$ & 1715.7 &	2.51E+02 & 1.47E-04	& 2.93E-04 \\
2076.8 & 7/2$^-$ & 1934.8 &	5.79E-02 & 2.45E-03	& 9.42E-03 \\
2583.8 & 3/2$^-$ & 2441.8 &	3.94E+03 & 1.48E-06	& 2.96E-06 \\
2597.1 & 5/2$^-$ & 2455.1 &	3.46E+01 & 3.37E-02	& 1.01E-01 \\
2636.8 & 7/2$^-$ & 2494.8 &	6.63E-01 & 5.06E-03	& 2.01E-02 \\
2937.0 & 3/2$^-$ & 2795.0 &	1.02E+04 & 1.72E-05	& 3.44E-05 \\
3041.7 & 5/2$^-$ & 2899.7 &	4.84E+01 & 1.07E-02	& 3.22E-02 \\
\bottomrule
\end{tabular*}
\vspace{0mm}
\end{center}
\vspace{0mm}

\subsubsection{Rate based on mirror structure}
Alternatively, we have estimated the $^{50}$Fe($p$,$\gamma$)$^{51}$Co resonant rate by using exactly the level energies, half-lives and single-particle
spectroscopic factors from the mirror nucleus $^{51}$Cr~\cite{bib:hua06}. A similar approach was utilized in Ref.~\cite{bib:hjj14}. Here, the gamma
widths ($\Gamma_\gamma$) of the unbound states in $^{51}$Co were estimated by the half-lives ($T_\mathrm{1/2}$) of the corresponding bound states
in the mirror $^{51}$Cr via $\Gamma_\gamma$=ln(2)$\times \hbar/T_\mathrm{1/2}$; the proton widths were calculated by Eq.~(6) in
Ref.~\cite{bib:wor94}, i.e.,
\begin{eqnarray}
\Gamma_{p}=\frac{3\hbar^2}{AR^2}P_{\ell}(E)C^2S_p.
\label{eq4}
\end{eqnarray}
with $R$=1.26$\times$(1+50$^{\frac{1}{3}}$) fm as the nuclear channel radius, and $C^2S_p$ the proton spectroscopic factor of the resonance. Here, we
assumed the proton spectroscopic factor in $^{51}$Co equal to the neutron spectroscopic factor in the $^{51}$Cr mirror, i.e., $C^2S_p=C^2S_n$.
The experimental neutron spectroscopic factors are adopted from the previous ($d$,$p$) transfer reactions~\cite{bib:rob68,bib:mac72,bib:cho77}.
Here, the spectroscopic factor is a model-dependent quantity (see e.g., Ref.~\cite{bib:wu14}), and its variation may change the proton width $\Gamma_{p}$
accordingly.

Here, we assumed a value of 0.001 for those high-spin states ($J^{\pi}$=9/2$^-$ and 11/2$^-$) listed in Table~\ref{tab3}, which have no available
experimental $C^2S$ values. Actually, for the states at $E_x$=1164.6, 2379.5, 2704.4, and 2767.3 keV, their contributions to the total resonant rate are
negligible with any values of $C^2S$$<$1; for the state at $E_x$=1480.1 keV, its contributions is negligible with any values under a condition of
$C^2S$$<$0.1, and we think this condition is appropriate for this $J^{\pi}$=11/2$^-$ state. For the $E_x$=2828.5 keV state ($J^{\pi}$=3/2$^-$),
similarly, its contribution can also be neglected with any values of $C^2S$$<$1. Therefore, the states discussed above play only a negligible role in the total
resonant rate in spite of assuming any $C^2S$ values. The exception is the $E_x$=2001.9 keV state ($J^{\pi}$=5/2$^-$) which plays an important role
in contributing the total rate. Its resonance strengths are about 80.5, 80.3, 78.3, and 63.1 meV for $C^2S$ values of 1.0, 0.1, 0.01, 0.001, respectively; such
variation in strengths cannot be regarded as substantial.

The resonant parameters derived above are listed in Table~\ref{tab3}, and the corresponding resonant rates (referred to as Res$^{mirror}$) in
Table~\ref{tab4}. The contributions of each resonance to the total resonant rate have been calculated, and the role for those important resonances
is shown in Fig.~\ref{fig1}. It shows that three key resonances (i.e., at $E_x$=749.1, 1352.7 and 2001.9 keV) dominate the total resonant rate in the
temperature region of 0.1$\sim$2.5 GK.

\begin{center}
\includegraphics[width=8cm]{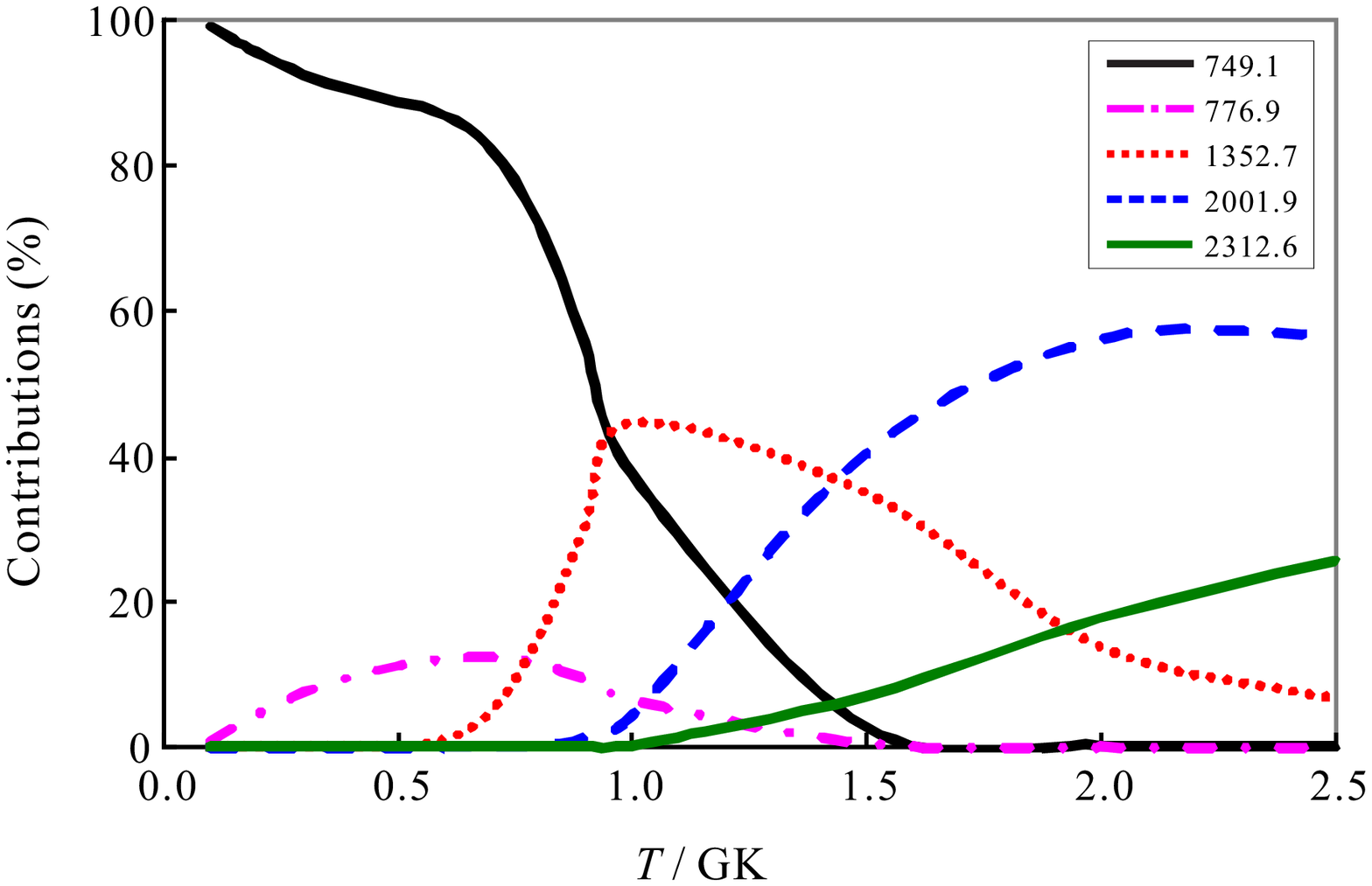}
\figcaption{\label{fig1} (color online) Percentage contribution of resonances to the total resonant rate. Here, 5 resonances (listed in Table~\ref{tab3}) which have
significant contribution ($>$10\%) are shown, with $E_x$ energies indicated in the legend.}
\end{center}

\end{multicols}
\begin{center}
\tabcaption{\label{tab3} Resonant parameters based on the nuclear structure in mirror $^{51}$Cr. The excitation energy $E_x$, spin-parity $J^{\pi}$, and
half-life $T_{1/2}$ are taken from Ref.~\cite{bib:hua06}. The spectroscopic factor values of $C^2S_p$ are the averaged ones from the previous ($d$,$p$)
experiments~\cite{bib:rob68,bib:mac72,bib:cho77}, except the assumed value of 0.001. }
\footnotesize
\begin{tabular*}{140mm}{cccccccc}
\toprule $E_x$ (keV) & $E_r^a$ (keV) & $J^{\pi}$ & $T_{1/2}$ (ps) & $C^2S_p$  & $\Gamma_\gamma^b$ (eV) & $\Gamma_p$ (eV) & $\omega\gamma$ (eV)  \\
\hline
\hphantom{0}749.1   & \hphantom{0}607.1  & 3/2$^-$                     & 3300	 &  0.36 	&  1.38E-07	&  2.48E-03	&  2.76E-07  \\
\hphantom{0}776.9   &	\hphantom{0}634.9  & 1/2$^-$                     &	6900	 &  0.29 	&  6.61E-08	&  4.12E-03	&  6.61E-08  \\
1164.6 & 1022.6 & 9/2$^-$                     & 	0.076 &  0.001	&  6.00E-03	&  9.64E-08	&  4.82E-07  \\
1352.7 & 1210.7 & 5/2$^-$                     & 	3.8	 &  0.17 	&  1.20E-04	&  1.57E-01	&  3.60E-04  \\
1480.1 & 1338.1 & 11/2$^-$                   &	0.55	 &  0.001	&  8.29E-04	&  3.56E-06	&  2.13E-05  \\
1557.3 & 1415.3 & 7/2$^-$                     & 	4.2	 &  0.09 	&  1.09E-04	&  4.78E-01	&  4.34E-04  \\
1899.2 & 1757.2 & 3/2$^-$                     & 	0.29	 &  0.16 	&  1.57E-03	&  5.47E+02	&  3.15E-03  \\
2001.9 & 1859.9 & 5/2$^-$                     & 	0.017 &  0.001	&  2.68E-02	&  9.74E-02	&  6.31E-02  \\
2312.6 & 2170.6 & 7/2$^-$                     & 	0.015 &  0.01 	&  3.04E-02	&  4.59E+00	&  1.21E-01  \\
2379.5 & 2237.5 & 9/2$^-$                     & 	0.31	 &  0.001	&  1.47E-03	&  1.42E-03	&  3.62E-03  \\
2704.4 & 2562.4 & (11/2$^-$)                   &	0.085 &  0.001	&  5.37E-03	&  5.83E-03	&  1.68E-02  \\
2762.6 & 2620.6 & 1/2$^+$                    &	0.071 &  0.02 	&  6.42E-03	&  2.53E+03	&  6.42E-03  \\
2767.3 & 2625.3 & 9/2$^-$                     & 	0.041 &  0.001	&  1.11E-02	&  7.45E-03	&  2.23E-02  \\
2828.5 & 2686.5 & 3/2$^-$                     & 	0.059 &  0.001	&  7.73E-03	&  9.29E+01	&  1.55E-02  \\
2890.2 & 2748.2 & 3/2$^-$                     & 	0.35	 &  0.10 	&  1.30E-03	&  1.03E+04	&  2.61E-03  \\
2911.0 & 2769.0 & (3/2$^-$,5/2,7/2$^-$)  & 0.03	 &  0.06 	&  1.52E-02	&  1.86E+02$^{f}$	&  4.56E-02  \\
2948.2 & 2806.2 & 5/2$^-$,7/2$^-$	       & 0.119	 &  0.04 	&  3.83E-03	&  1.63E+02	&  1.15E-02  \\
\bottomrule
\end{tabular*}
\vspace{0mm}

$^a$: Calculated by $E_r=E_x-S_p^\mathrm{IMP}$ with $S_p^\mathrm{IMP}$=142 keV~\cite{bib:shu14}.
$^b$: Calculated by $\Gamma_\gamma$=ln(2)$\times \hbar$/$T_\mathrm{1/2}$.
$^f$: Assuming an $\ell$=3 transfer.
\end{center}

\begin{multicols}{2}

\subsection{DC rate}
\label{sec:DC}
The nonresonant direct-capture (DC) rate can be estimated by the following expression~\cite{bib:rol88},
\end{multicols}
\ruleup
\begin{equation}
\label{eq5}
N_A\langle \sigma v \rangle_\mathrm{DC}=7.83 \times 10^{9} \left( \frac{Z_T}{A} \right)^{1/3}T_9^{-2/3}S^\mathrm{eff}_\mathrm{DC}(E_0)\times \mathrm{exp} \left[ -4.249 \left (\frac{Z_T^2 A}{T_9} \right)^{1/3} \right] \quad [\mathrm{cm^3s^{-1}mol^{-1}}],
\end{equation}
\begin{multicols}{2}
\noindent
with $Z_T$ being the atomic number of $^{50}$Fe. The effective astrophysical $S$-factor at the Gamow energy $E_0$, i.e., $S^\mathrm{eff}_\mathrm{DC}(E_0)$,
is usually approximated by~\cite{bib:fis01,bib:rol88},
\begin{eqnarray}
S^\mathrm{eff}_\mathrm{DC}(E_0)=S(0)\left( 1+\frac{5}{12\tau} \right),
\label{eq6}
\end{eqnarray}
where $S$(0) is the $S$-factor at zero energy. It should be noted that the direct-capture rate expressed by Eq.~(15) in Ref.~\cite{bib:fis01} should be
multiplied by a factor of 2.

In this work, we have calculated the $^{50}$Fe($p$,$\gamma$)$^{51}$Co DC $S$-factors with a \texttt{RADCAP} code~\cite{bib:ber03}. The
Woods-Saxon nuclear potential (central + spin orbit) and a Coulomb potential of uniform-charge distribution were utilized in the calculation. The nuclear
central potential $V_0$ were determined by matching the bound-state energies. The optical-potential parameters~\cite{bib:fis01} are
$R_0=R_\mathrm{s.o.}=R_C=1.25\times(1+A^\frac{1}{3})$~fm, $a_0=a_\mathrm{s.o.}=0.6$~fm, with a depth of spin-orbit potential of
$V_\mathrm{s.o.}=-10$ MeV. Here, $R_0$, $R_\mathrm{s.o.}$ and $R_C$ are radii of central potential, spin-orbit potential and Coulomb potential,
respectively; $a_0$ and $a_\mathrm{s.o.}$ are the corresponding diffuseness in central and spin-orbit potentials, respectively. We have reproduced the
previous value of $S$(0)=0.1313 [MeV b] with a spectroscopic factor of $C^2S$=0.22, by using the same optical-model parameters and an $S_p$ value of
90 keV as in Ref.~\cite{bib:fis01}. This spectroscopic factor can be found in the Ref.~\cite{bib:rob68}.

Here, we adopted an averaged spectroscopic factor of $C^2S$=0.29 for the ground-state capture~\cite{bib:rob68,bib:mac72,bib:cho77}, together with the
new experimental value of $S_p^\mathrm{IMP}$=142(77) keV and the same optical-model parameters of Ref.~\cite{bib:fis01}. A value of $S$(0)=0.1896
[MeV b] is obtained for the DC capture. The presently calculated $S_\mathrm{DC}$ factors can be well parameterized in a Taylor-series
form~\cite{bib:rol88} of $S_\mathrm{DC}(E)=S(0)+\dot{S}(0)E+\frac{1}{2}\ddot{S}(0)E^2$, where $S$ factor are in units of [MeV b] and $E$ in MeV.
The fitted parameters are $S(0)$=0.1896 [MeV b], $\dot{S}(0)$=5.897$\times$10$^{-2}$ [MeV$^{-1}$] and $\ddot{S}(0)$=4.787$\times$10$^{-2}$ [MeV$^{-2}$],
respectively. According to the textbook~\cite{bib:rol88}, the effective astrophysical $S$-factor at the Gamow energy $E_0$ in the above Eq.~(\ref{eq5})
can be expressed as the well-known formulism~\cite{bib:rol88},
\end{multicols}
\ruleup
\begin{equation}
\label{eq7}
S^\mathrm{eff}_\mathrm{DC}(E_0)=S(0)\left[1+\frac{5}{12\tau}+\frac{\dot{S}(0)}{S(0)}\left(E_0+\frac{35}{36}kT \right)+\frac{1}{2}\frac{\ddot{S}(0)}{S(0)}\left(E_0^2+\frac{89}{36}E_0kT \right) \right].
\end{equation}
\begin{multicols}{2}
\noindent

The DC reaction rates calculated with the approximated Eq.~(\ref{eq6}) are compared to those with the more precise Eq.~(\ref{eq7}), and we find that
the latter are larger than the former by about factors of 1.1, 2.5, and 7.0 at 0.1, 3, and 10 GK, respectively; the latter are consistent very well with the
numerical integration method by using an EXP2RATE code~\cite{bib:exp2}. Thus, we have calculated the DC rates with Eqs.~(\ref{eq5})~\&~(\ref{eq7})
as listed in Table~\ref{tab4}.

In addition, the parameter dependence on $S_\mathrm{DC}(E)$ has been studied for this reaction. The sensitivities are about: 80\% on
$R_0$ (4.64$\sim$5.86 fm, i.e., 1.25$\times$(1+50)$^\frac{1}{3}$=4.64 fm~\cite{bib:hua10}, 1.25$\times$(1+50$^\frac{1}{3}$)=5.86 fm~\cite{bib:fis01}),
12\% on $V_\mathrm{s.o.}$ (0$\sim$-10 MeV~\cite{bib:hua10}), 10\% on $S_p$ error ($\pm$77 keV), and 8\% on $a$ (0.55~0.65 fm), respectively.
The uncertainties of the derived S factors and DC rates are about a factor of 3.

Comparing the resonant rates and DC rate listed in Table~\ref{tab4}, it shows that the DC contribution dominate the total rate below 0.15 GK, beyond which
the resonant capture makes the overwhelming contribution. Our result is significantly different from the previous conclusion~\cite{bib:fis01}.

\begin{center}
\tabcaption{\label{tab4} Presently calculated resonant and DC rates for the $^{50}$Fe($p$,$\gamma$)$^{51}$Co reaction, in units of
cm$^3$s$^{-1}$mol$^{-1}$.}
\footnotesize
\begin{tabular*}{85mm}{c@{\extracolsep{\fill}}cccc}
\toprule $T_9$  & DC  & Res$^{shell}$  & Res$^{mirror}$ \\
\hline
0.10 &	4.89E-25 &	8.26E-34 &	3.49E-31 \\
0.15 &	9.22E-21 &	2.46E-22 &	3.06E-21 \\
0.20 &	4.58E-18 &	1.21E-16 &	2.55E-16 \\
0.30 &	1.10E-14 &	5.22E-11 &	1.80E-11 \\
0.40 &	1.49E-12 &	3.06E-08 &	4.25E-09 \\
0.50 &	4.89E-11 &	1.31E-06 &	1.05E-07 \\
0.60 &	7.05E-10 &	1.53E-05 &	8.54E-07 \\
0.70 &	5.94E-09 &	8.54E-05 &	3.82E-06 \\
0.80 &	3.46E-08 &	3.02E-04 &	1.26E-05 \\
0.90 &	1.54E-07 &	7.92E-04 &	3.64E-05 \\
1.00 &	5.56E-07 &	1.68E-03 &	1.00E-04 \\
1.50 &	5.31E-05 &	1.46E-02 &	7.52E-03 \\
2.00 &	9.66E-04 &	4.99E-02 &	1.28E-01 \\
2.50 &	7.73E-03 &	1.72E-01 &	7.85E-01 \\
\bottomrule
\end{tabular*}
\vspace{0mm}
\end{center}
\vspace{0mm}

\subsection{Total reaction rates}
The total reaction rate of $^{50}$Fe($p$,$\gamma$)$^{51}$Cr has been calculated by simply summing up the resonant and DC rates as discussed above.
Two sets of total rates, referred to as \emph{Shell} and \emph{Mirror}, are tabulated in Table~\ref{tab5}. The present \emph{Mirror} rate can be well
parameterized by the standard format of~\cite{bib:rau00}:
\end{multicols}
\ruleup
\begin{eqnarray}
\label{eq8}
N_A\langle\sigma v\rangle &=& \mathrm{exp}(687.603-\frac{9.111}{T_9}+\frac{575.394}{T_9^{1/3}}-1378.820T_9^{1/3}+128.398T_9-13.263T_9^{5/3}+549.246\ln{T_9}) \nonumber \\[1mm]
&&
+ \mathrm{exp}(578.307-\frac{16.326}{T_9}+\frac{541.104}{T_9^{1/3}}-1223.680T_9^{1/3}+123.472T_9-12.881T_9^{5/3}+475.982\ln{T_9})
\end{eqnarray}
\begin{multicols}{2}
\noindent
with fitting error of less than 0.6\% in 0.1--2.5 GK; the present \emph{Shell} rate can be expressed as,
\end{multicols}
\ruleup
\begin{eqnarray}
\label{eq8}
N_A\langle\sigma v\rangle &=& \mathrm{exp}(2129.450+\frac{20.158}{T_9}-\frac{1028.750}{T_9^{1/3}}-2426.040T_9^{1/3}+2276.410T_9-1051.510T_9^{5/3}-325.917\ln{T_9}) \nonumber \\[1mm]
&&
+ \mathrm{exp}(-254.347-\frac{5.474}{T_9}-\frac{179.555}{T_9^{1/3}}+485.685T_9^{1/3}-61.715T_9+9.015T_9^{5/3}-173.845\ln{T_9})
\end{eqnarray}
\begin{multicols}{2}
\noindent
with fitting error of less than 0.5\% in 0.1--2.5 GK. We emphasize that the above fits are only valid within the stated errors over the temperature range of
0.1--2.5 GK. Above 2.5 GK, one may, for example, match our rates to those statistical model calculations.

The comparison between different rates relative to the present \emph{Mirror} rate is shown in Fig.~\ref{fig2}. The differences are explained below:
(1) the statistical-model rates
(\emph{rath, ths8, thra}) are about 1$\sim$6 orders of magnitude larger, which demonstrates that the statistical-model is not ideally applicable for this
reaction mainly owing to the low density of low-lying excited states in $^{51}$Co.
(2) the \emph{laur} rate based on the mirror information of $^{51}$Cr is
about a factor of 30$\sim$650 times larger at temperature $>$0.15 GK, mainly because of two factors: one is the different $S_p$ values utilized, another is
the inappropriate approximation ($\omega\gamma$$\approx$$\omega\Gamma_\gamma$) made in the previous work as discussed in Sec.2.2; below $\sim$0.15 GK,
\emph{laur} rate decreases because that its DC contribution was neglected~\cite{bib:wor94}.
(3) In the temperature region of 1$\sim$2.5 GK, the \emph{nfis} rate is almost the closest one (with deviation less than $\sim$50\%) to the \emph{Mirror} rate;
below 1 GK, \emph{nfis} is about 1$\sim$3 orders magnitude smaller mainly because of a relatively larger value of $S_p$=0.355 MeV utilized.
(4) The present \emph{Shell} rate agree with the \emph{Mirror} one within a factor of up to about 10.

\begin{center}
\includegraphics[width=8cm]{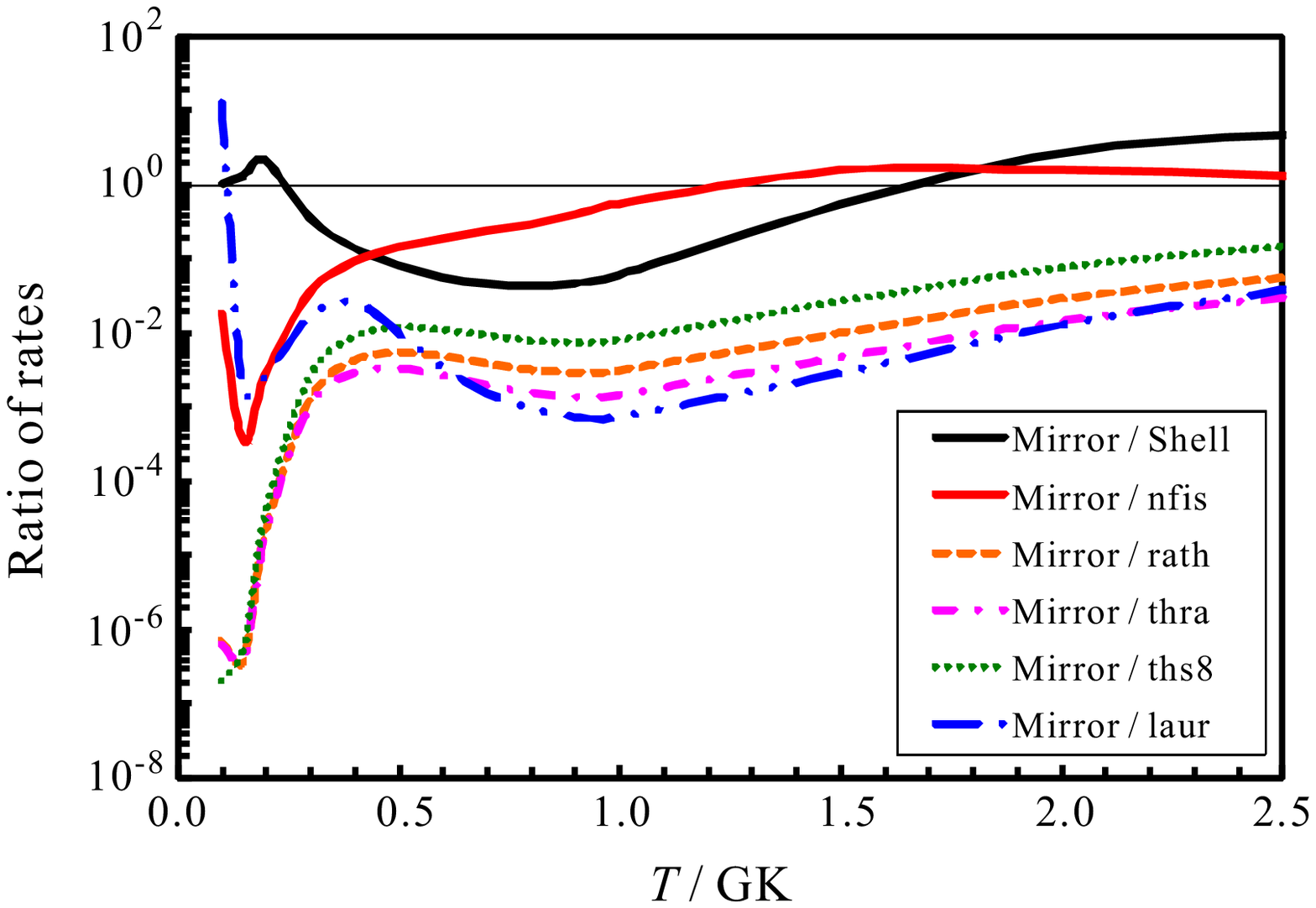}
\figcaption{\label{fig2} (color online) Comparison between different reaction rates relative to the present \emph{Mirror} rate. The reference of unity is indicated by
a solid line.}
\end{center}

\end{multicols}
\begin{center}
\tabcaption{\label{tab5} Thermonuclear rates of $^{50}$Fe($p$,$\gamma$)$^{51}$Co. The adopted $S_p$ values are listed in the parentheses.}
\footnotesize
\begin{tabular*}{155mm}{ccccccccc}
\toprule & \multicolumn{2}{c}{Present rates} & & \multicolumn{5}{c}{JINA REACLIB rates} \\
\cline{2-3} \cline{5-9}\\
      &  \emph{Mirror}   & \emph{Shell} &  & \emph{nfis}  & \emph{rath} & \emph{thra} & \emph{ths8}  & \emph{laur} \\
$T_9$ &  (0.142 MeV) & (0.142 MeV) &  & (0.355 MeV)  & (1.369 MeV) & (1.659 MeV) & (0.355 MeV)  & (0.24 MeV) \\
\hline
0.10  &	4.89E-25 &	4.89E-25 &	& 2.67E-23 &	7.28E-19 &	7.46E-19 &	2.42E-18 &	3.85E-26 \\
0.15  &	1.23E-20 &	9.47E-21 &	& 3.47E-17 &	3.37E-14 &	3.14E-14 &	2.35E-14 &	7.99E-18 \\
0.20  &	2.60E-16 &	1.26E-16 &	& 9.63E-14 &	1.33E-11 &	1.31E-11 &	6.85E-12 &	1.02E-13 \\
0.30  &	1.80E-11 &	5.22E-11 &	& 5.08E-10 &	1.47E-08 &	1.73E-08 &	7.02E-09 &	1.12E-09 \\
0.40  &	4.25E-09 &	3.06E-08 &	& 4.67E-08 &	1.03E-06 &	1.44E-06 &	4.82E-07 &	1.63E-07 \\
0.50  &	1.05E-07 &	1.31E-06 &	& 7.28E-07 &	1.99E-05 &	3.15E-05 &	8.94E-06 &	1.05E-05 \\
0.60  &	8.55E-07 &	1.53E-05 &	& 4.51E-06 &	1.81E-04 &	3.16E-04 &	7.79E-05 &	2.51E-04 \\
0.70  &	3.83E-06 &	8.54E-05 &	& 1.65E-05 &	1.02E-03 &	1.90E-03 &	4.19E-04 &	2.47E-03 \\
0.80  &	1.26E-05 &	3.02E-04 &	& 4.35E-05 &	4.08E-03 &	7.97E-03 &	1.63E-03 &	1.35E-02 \\
0.90  &	3.65E-05 &	7.92E-04 &	& 9.41E-05 &	1.28E-02 &	2.58E-02 &	4.97E-03 &	4.99E-02 \\
1.00  &	1.01E-04 &	1.68E-03 &	& 1.82E-04 &	3.34E-02 &	6.86E-02 &	1.27E-02 &	1.40E-01 \\
1.50  &	7.57E-03 &	1.47E-02 &	& 4.90E-03 &	7.81E-01 &	1.60E+00 &	2.90E-01 &	2.67E+00 \\
2.00  &	1.29E-01 &	5.08E-02 &	& 8.28E-02 &	4.59E+00 &	8.79E+00 &	1.73E+00 &	1.03E+01 \\
2.50  &	7.93E-01 &	1.79E-01 &	& 5.99E-01 &	1.48E+01 &	2.65E+01 &	5.55E+00 &	2.16E+01 \\
\bottomrule
\end{tabular*}
\vspace{2mm}
\end{center}
\begin{multicols}{2}

\section{Summary}
The thermonuclear rate (including direct-capture (DC) and resonant contribution) of the $^{50}$Fe($p$,$\gamma$)$^{51}$Co reaction has been recalculated
by utilizing the recent precise proton separation energy of $S_p$($^{51}$Co)=142$\pm$77 keV measured at the HIRFL-CSR facility in Lanzhou, China.
Here, the resonant rates have been calculated in two ways: one is to revise the previous shell-model results with this new $S_p$ value (i.e., \emph{Shell} rate),
another is to rely on the mirror nuclear structure of $^{51}$Cr (i.e., \emph{Mirror} rate). Our new rates deviate significantly from those available in the
literature. We conclude that statistical model calculations are not ideally applicable for this reaction primarily because of the low density of low-lying excited
states in $^{51}$Co. Thus, we recommend that the present new \emph{Mirror} rate should be incorporated in the future astrophysical network
calculations, since it is based on more solid experimental background. The astrophysical impact of our new rates in Type I x-ray burst calculations is now under
progress, which is beyond the scope of this work.


\end{multicols}

\vspace{2mm}
\centerline{\rule{80mm}{0.1pt}}
\vspace{2mm}

\begin{multicols}{2}

\end{multicols}

\clearpage
\end{CJK*}
\end{document}